\documentclass[12pt]{article}
\begin{document}

\begin{center}
{\large \bf Exact Solutions of a Non-Polynomially Nonlinear Schrodinger Equation}
\end{center}
\vspace{0.1in}

\begin{center}

{R. Parwani\footnote{Email: parwani@nus.edu.sg} and H. S. Tan}

\vspace{0.3in}

{Department of Physics,\\}
{National University of Singapore,\\}
{Kent Ridge,\\}
{ Singapore.}

\vspace{0.3in}
8 May 2006;
Revised 3 September 2006; Revised 1 Nov 2006
\end{center}
\vspace{0.1in}
\begin{abstract}

A nonlinear generalisation of Schrodinger's equation had previously been obtained using information-theoretic arguments.  The nonlinearities in that equation were of a nonpolynomial form, equivalent to the occurence of higher-derivative nonlinear terms at all orders. Here we construct some exact solutions to that equation in $1+1$ dimensions. On the half-line,  the solutions resemble (exponentially damped) Bloch waves eventhough no external periodic potential is included. The solutions are nonperturbative as they do not reduce to solutions of the linear theory in the limit that the nonlinearity parameter vanishes. An intriguing feature of the solutions is their infinite degeneracy: for a given energy, there exists a very large arbitrariness in the normalisable wavefunctions. We also consider solutions to a $q$-deformed version of the nonlinear equation and discuss a natural discretisation implied by the nonpolynomiality.  Finally, we contrast the properties of our solutions with other solutions of nonlinear Schrodinger equations in the literature and suggest some possible applications of our results in the domains of low-energy and high-energy physics.

\end{abstract}

\vspace{0.5in}

\section{Background}

Nonlinear extensions of Schrodinger's equation are often used as effective theories in domains such as optics \cite{optics} and condensed matter physics \cite{bose}, the prototype being the Gross-Pitaevskii equation \cite{sulem}. However, nonlinear Schrodinger equations have also been used to probe departures from exact quantum linearity \cite{probe} with several experiments setting upper bounds on some of the proposed nonlinearities \cite{bdds}. The literature on theoretical investigations into nonlinear Schrodinger equations is immense, see for example the citations in \cite{lit}.

In Ref.\cite{me-ann} a particular nonlinear extension of Schrodinger's equation was motivated by information theory arguments \cite{fried1,P1} similar to those used in statistical mechanics \cite{Jay}. For a single particle in one space dimension, the obtained equation was 
\begin{equation}
i \hbar {\partial \psi \over \partial t} = - {{\hbar}^2 \over 2m} {\partial^2 \psi \over \partial x^2} + V(x) \psi + 
F(p) \psi \, , \label{nsch1}
\end{equation}
with
\begin{eqnarray} 
F(p) &=& {\cal{E}} \left[ \ln {p(x) \over p(x+L)}  \ + 1 \ - {p(x-L) \over p(x)} \right] + {{\hbar}^2 \over 2m}  {1 \over \sqrt{p}} {\partial^2 \sqrt{p} \over \partial x^2} \label{fip}
\end{eqnarray}
and $p(x) = \psi^{\star}(x) \psi(x)$. The energy scale ${\cal{E}}$ is related to the length scale $L$ through the relation 
\begin{equation}
{\cal{E}} L^2 = {{\hbar}^2  \over 4 m } \, , \label{uncert}
\end{equation}
to ensure that the leading term in the small $L$ expansion of (\ref{nsch1}) yields the usual linear Schrodinger equation.
Note that if ${\cal{E}}$ is finite, then $L$  cannot be zero since $\hbar$ is non-vanishing.  

The nonlinear Schrodinger equation (\ref{nsch1}) shares a number of properties with the linear Schrodinger equation. Of importance is that $p(x)=\psi^{*} \psi$ still satisfies the usual continuity equation and thus has a sensible interpretation as probability. Furthermore,  equation (\ref{nsch1}) still admits plane wave solutions $\psi = e^{ikx-i\omega t}$ and more generally, if  $\psi$ is any solution of the equation, then so is $\lambda \psi$ for an arbitrary  constant $\lambda$. 

As written, the nonlinear term $F (p)$ is not invariant under the parity transformation $x \to -x$. A parity-invariant equation is easily obtained by using a symmetrised form of Eq.(\ref{nsch1}) involving both postive and negative values of $L$, 
\begin{equation}
F(p) \to {1 \over 2} ( F^{+}(p) + F^{-}(p) ) \label{symmp}
\end{equation}
where $F^{+}$ is the same as $F$ but $F^{-}$ uses $-L$ instead of $L$. 
There is no problem in extending the nonlinear Schrodinger equation  to more than one particle or higher dimensions, but in higher dimensions the equation is not rotationally invariant, leading to the suggestion \cite{me-ann} that spacetime symmetries might be linked to quantum linearity, with $L$ possibly linked to gravitational effects \cite{P3}. 

Given a nonlinear extension, it is natural to investigate whether any novel effects arise, such as the occurence of solitary waves, and to seek their experimental confirmation. In \cite{me-ann} some preliminary phenomenological consequences of the nonlinearity were suggested based mainly on qualitative and heuristic arguments. It was also noted in \cite{me-ann} that the equation had a unusual exact solution:  
Consider the free case,$V=0$, of equation (\ref{nsch1}) for a stationary state $\psi = e^{-iEt/ \hbar} \phi(x)$ with $\phi$ real. Then substitution yields  
\begin{eqnarray}
E &=& - {{\hbar}^2 \over 2m \phi } {\partial^2 \phi \over \partial x^2} + {\cal{E}} \left[ \ln {p(x) \over p(x+L)}  \ + 1  - {p(x-L) \over p(x)} \right] + {{\hbar}^2 \over 2m \sqrt{p} }  {\partial^2 \sqrt{p} \over \partial x^2} \label{zfirst}\\
&=&{\cal{E}} \left[ \ln {p(x) \over p(x+L)}  \ + 1 \ - {p(x-L) \over p(x)} \right] \, . \label{1dex}
\end{eqnarray}
Clearly for {\it any} $\phi(x)$ that is periodic in the nonlinearity scale $L$, 
\begin{equation}
\phi(x+L) = \phi(x),
\end{equation}
the right-hand side of the eigenvalue equation (\ref{1dex}) vanishes and so $E=0$.  These zero-energy periodic solutions exist only for the exact equation (\ref{1dex}) but not  for perturbative approximations of the  equation obtained by expanding to some power in $L$. For example, the perturbative approximation to (\ref{nsch1}) using 
\begin{equation}
F(p) = {\hbar^2 L \over 4m} \left[ - { (p')^3 \over 3 p^3} + {p' p'' \over 2 p^2} \right] + O(L^2) \, , \label{f1est}
\end{equation} 
where the prime denotes derivative with respect to $x$, does not support the class of periodic solutions of the exact equation mentioned above. 

Our purpose in this paper is to construct an even larger class of exact, localised, solutions for which the abovementioned zero-energy periodic solutions are a limiting case. We begin our construction in Sect.(2) for a single particle confined to a box.  A continuum spectrum of nonperturbative solutions is obtained which  is not bounded below, even when a regularised version \cite{me-ann} of eq.(\ref{nsch1}) is used. However when the equation is reconsidered  on the half-line, then the  spectrum of the solutions is bounded from below. In Sect.(3) we study a $q$-deformed version \cite{me-ann} of the equation (\ref{nsch1}) and again construct a continuum set of nonperturbative solutions whose spectrum is bounded from below for some values of the parameter $q$. 

Equation (\ref{nsch1}) suggests a obvious discretisation which we discuss in Sect.(4). Some approximate analytical solutions representing localised states on the infinite line are also dicussed in Sect.(4): They might be useful seeds for further work. A discussion of potential applications and future directions  appears in the final section.

\section{Regularisation and Solutions}
As the nonlinear term in (\ref{nsch1}) might have singularities where $p(x)$ vanishes, it is convenient to consider a regularised version of that expression. Defining 
\begin{eqnarray}
p_{\pm}(x) &=& p(x \pm \eta L) \,  ,
\end{eqnarray}
where the dimensionless parameter $\eta$ takes values $0 < \eta \le 1$, the regularised  potential  
\begin{equation}
Q_{1}^{\eta} = { {\cal{E}}_i  \over \eta^4}  \left[ \ln {p \over (1-\eta) p + \eta p_{+} } + 1 - {(1-\eta) p \over (1-\eta) p + \eta p_{+}} - {\eta p_{-} \over (1-\eta) p_{-} + \eta p} \right] \,  \label{Q2}
\end{equation}
was constructed in \cite{me-ann}, and is  to be used in (\ref{fip}) instead of the first term there. With this regularisation there are no singularities in the equation of motion. Formally, the $\eta$ parameter can be also used to interpolate between the fully nonlinear theory at $\eta =1$ and the usual linear quantum mechanics at $\eta =0$. 

We wish to find exact stationary states of the regularised nonlinear equation. Notice that if
\begin{equation}
p(x \pm \eta L) = g(L) p(x) \label{prop}
\end{equation}
for some function $g(L)$, then the term (\ref{Q2}) simplifies greatly, thus facilitating a solution of the nonlinear equation. Condition (\ref{prop}) is equivalent to the eigenvalue equation for the translation operator, 
\begin{equation}
\exp{ (\pm \eta L \partial_{x} )} p(x) = g(L) p(x) \, ,
\end{equation}
which is satisfied by a $p(x)$ of exponential form. More generally, just as in the example for (\ref{1dex}), one may multiply that exponential solution  by a function periodic in $ \eta L$ and still obtain the simplification of (\ref{Q2}). Thus the {\it ansatz} we choose for our class of solutions is 
\begin{equation}
\psi(x,t) = C \exp{(-\kappa x)} \ \alpha(x) \ \exp{(-iEt/ \hbar)} \label{ants}
\end{equation}
where $C$ is the normalisation, $\kappa$ a real parameter and  $\alpha(x)$ is {\it any} periodic function satisfying 
\begin{equation}
\alpha(x \pm \eta L) = \alpha (x) \, . \label{arb}
\end{equation}

For the ansatz (\ref{ants})  the regularised nonlinear Schrodinger equation reduces to the following relation between the energy, $E$, and the other parameters,
\begin{equation}
(1- { E \eta^4 \over  {{\cal E}} } ) = \ln [ 1 + \eta (\gamma -1) ] + {1  \over 1 + \eta (\gamma-1) } \label{energy}
\end{equation}
where 
\begin{equation}
\gamma \equiv \exp{(-2\kappa \eta L)} \, . \label{gam}
\end{equation}
For definiteness one may take,
\begin{equation}
\alpha(x) = \sin( {2 \pi x \over \eta L} ) \, , \label{deff}
\end{equation} 
then on the domain $0 \le x \le N \eta L$ with $N$ an integer, the normalisation factor is
\begin{equation}
C^{-2} = { (1- \gamma^N) \over 4 \kappa}  {16 \pi^2 \over (\ln \gamma)^2 + 16 \pi^2} \, .
\end{equation}  
With the choice (\ref{deff}) the wavefunction (\ref{ants}) vanishes at the walls of the infinite well as required on physical grounds. Note that the equations (\ref{nsch1}, \ref{Q2}) require us to define the wavefunction simultaneously at points $x, x\pm \eta L$ for each $x$ in the domain $[0,a]$. This means that we must also give values to the wavefunction for some points outside the physical domain. The way we do this is to use (\ref{ants}) throughout: This convention\footnote{Of course one could adopt other prescriptions, such as periodic boundary conditions. If one imposed periodic boundary conditions on (\ref{ants}), then $\kappa =0$ and one obtains the zero energy solution first found in Ref.\cite{me-ann}. That solution is still non-perturbative and highly degenerate, see (\ref{arb}).} preserves not only the continuity of the wavefunction but ensures the validity of the eigenvalue equation (\ref{energy}) for all points in $[0,a]$.

Then, since on the finite $x-$domain the solution (\ref{ants},\ref{deff}) is normalisable for any $ -\infty \le \kappa \le \infty $, it  means that for any given $\eta$ the variable $\gamma$ can take values in the range $ 0 \le \gamma \le \infty$. Define $\theta  = 1 + \eta(\gamma -1)$. Then on the finite $x-$domain, $(1- \eta) \le \theta \le \infty$ and so the function 
\begin{equation}
y = \ln \theta + {1 \over \theta}
\end{equation}   
is unbounded above, with a global minimum at $\theta =1$ (corresponding to $\gamma = 1$). Thus the energy given by (\ref{energy}) would be unbounded from below. 

Since it is necessary to have the energy bounded from below to ensure an energetically  stable ground state, 
one clearly has to restrict the range of $\theta$ in some natural way. This can be achieved by taking $N \to \infty$ so that one is considering dynamics on the semi-infinite line $0 \le x \le \infty$. Normalisability of the solutions now requires $\kappa > 0$ in (\ref{gam}), and so $ 0 \le \gamma < 1$ and $(1- \eta) \le \theta  < 1$. Correspondingly, the energy is now restricted to 
\begin{equation}
0 > E \ge { {\cal{E}} \over \eta^4} \left( 1- \ln (1- \eta) - {1 \over (1 -\eta) } \right) \, .
\end{equation}

As mentioned above, in the limit $\eta \to 0$ the nonlinear equation reduces to the ususal linear equation. In the same limit, the energy (\ref{energy}) becomes,
\begin{equation}
 E  \to -{{\hbar}^2  \kappa^2 \over 2 m } \, ,
\end{equation}
where (\ref{uncert}) has been used. However as (\ref{deff}) indicates, on the semi-infinite domain the wavefunction itself (\ref{ants}) does not reduce to that of the linear theory but instead undergoes infinitely fine oscillations. Thus even an infinitesimal  nonlinearity will leave its signature on the quantum ground state through nonperturbative effects: such possibilities are commonly studied in field theory contexts but we see them here at the quantum mechanical level. 

The above discussion has been for the parity-violating case of the nonlinear equation. If one considered the parity symmetric version of the equation using (\ref{symmp}), then it is easy to check that even for dynamics on a semi-infinite region of the $x$-axis the energy would be unbounded below.

\section{A Q-deformation}

Another nonlinear Schrodinger equation that was constructed in \cite{me-ann} used a different information measure 
\cite{THC}, based on the deformed logarithm  defined by 
\begin{equation}
\ln_{q} y = {y^{q-1} -1 \over q-1} \, 
\end{equation}
where the usual logarithm is recovered as the real parameter $q \to 1$. The net result is a nonlinear Schrodinger equation as in (\ref{nsch1}) but with the first term in (\ref{fip}) replaced by the a q-deformed quantum potential
\begin{eqnarray}
Q_{NL}^{(q)} &=& { {\cal{E}} \over q } \left[ \ln_q {p \over p_{+}} + \left({p \over p_{+}}\right)^{q-1} - \left({p_{-} \over p}\right)^q \right] \, . 
\end{eqnarray}
with the other symbols having their previous meanings and $q > 0$ in order to recover the usual linear Schrodinger equation in the small $L$ limit.  

We can construct solutions to the q-deformed nonlinear Schrodinger equation using the same {\it ansatz} (\ref{ants}) as before. The only thing that changes is the energy eigenvalue relation which now reads 
\begin{equation}
{ q E \over {\cal E} } + { 1 \over q-1} = \lambda ^{q-1} \left( { q \over q-1} - \lambda \right)  \label{eng2}
\end{equation}
where we have defined 
\begin{equation}
\lambda \equiv \exp{ (2 \kappa L)} \, .
\end{equation} 
Again one finds that unless $\kappa$ is restricted in some natural way, the energy is unbounded from below. For dynamics on the half-line $ -\infty \le x \le 0$, normalisability of solutions requires $\kappa < 0$ and so  $0 \le \lambda < 1$. Then one finds that the energy as given by (\ref{eng2}) is bounded from below (and above) if one chooses $q > 1$.

\section{ Approximate Solutions and Discretisation}

The ansatz (\ref{ants}) does not permit localised solutions on the whole x-axis and it appears difficult to construct exact solutions with some other ansatz. However, reasonably good approximate solutions can be constructed by patching the existing solutions as follows. Let
\begin{equation}
\psi(x,t) = C' \exp{(-iEt/ \hbar)} \ \sin( {2 \pi x \over \eta L} ) \ \left( \theta(x) \phi_{+}(x) + \theta(-x) \phi_{-}(x) \right) \, , \label{patch}
\end{equation}  
where $\theta(x)$ is the usual step function, $C'$ a normalisation factor and 
\begin{equation}
\phi_{\pm } (x) = \exp{(\mp \kappa_{\pm}  x)} \, ,  
\end{equation}
with $\kappa_{\pm} >0$. This normalisable and smooth wavefunction clearly satisfies the regularised nonlinear equation of Sect.(2) {\it outside} the small interval $-\eta L \le x \le \eta L$ when only one patch of the wavefunction contributes. The values of $\kappa_{+}$ and $\kappa_{-}$ are related to each other and to the energy $E$ through a relation similar to eqs.(\ref{energy},\ref{gam}).  In the small interval around the origin eq.(\ref{patch}) is not a solution so one requires a modification there. Thus the ansatz (\ref{patch}) might be adopted as a trial for numerical searches of localised solutions, though we do not know if it would be useful in practice.  

In low-energy physics, nonlinear differential equations often arise as continuum approximations to difference equations defining, for example, dynamics on a lattice of atoms, see for example \cite{nonlocal1}. On the other hand, in high-energy physics, although spacetime is usually taken to be continuous, there have been numerous suggestions \cite{Garay} that at a more fundamental level some form of discreteness should arise. Thus for both of those reasons, we look at discretisations of the nonlinear Schrodinger equations studied above. Let us focus on stationary states described by the basic equation (\ref{1dex}).  It is remarkable that that equation has no explicit space derivatives, so the usual ambiguities of how to discretise a differential equation do not arise. In fact the structure of the nonpolynomial equation (\ref{1dex}), which contains the function $p(x)$ and its shifted values $p(x \pm L)$, suggests a natural discretisation: Simply set $L=1$, $p(x) \to p_n, \, p(x \pm L) \to p_{n \pm 1}$, with $n$ an integer, leading to a nonlinear difference equation for the energy eigenvalue. 

The above discretisation extends obviously to the regularised and q-deformed equations. The discrete forms of the equations might be more convenient for discovering new solutions in future investigations.

\section{Discussion}  

The nonlinear Schrodinger equation (\ref{nsch1}), that was motivated by information theoretic arguments in \cite{me-ann}, has a fundamentally different structure from other nonlinear extensions of Schrodinger's equation that we are aware of. Indeed, the nonlinear equation we have studied has the form of a differential-difference equation with points at $x, x\pm \eta L$ simultaneously required in its definition. When expanded in $L$, the equation gives rise to higher derivatives of nonlinear terms at all orders. 

Furthermore, the nonlinearities in (\ref{nsch1}) are of a nonpolynomial form.  
It is interesting to note that nonpolynomial nonlinearities have been considered in a physical context 
\cite{nonpoly} but no higher order derivatives occured there. In a different context, nonlinear Schrodinger equations with derivatives at all orders have been considered in \cite{nonlocal1}, but that nonlocality was only in the linear part of the equation. Thus the equation (\ref{nsch1}) combines  features seen in separate equations in the literature, and perhaps it might find a physical application in those contexts.  

The nonlinear equation (\ref{nsch1}) has given rise to interesting solutions. Even in a bounded region of space, the energies of our class of solutions form a continuous spectrum and furthermore the wavefunctions are highly degenerate, as indicated by the arbitrariness of the periodic function $\alpha(x)$. These features appear to be very different from various other exact solutions and discretisations of nonlinear Schrodinger equations studied in the literature, see for example 
\cite{nonlocal1, nonpoly, nonlink, nonlin-others} and references therein.

From a high-energy physics perspective, as originally envisaged in \cite{me-ann}, the equation might be relevant for probing possible new physics at short distances. In this regard we note that higher dimensional, separable versions of the equation are available \cite{me-ann}. If one considers extra-dimensional scenarios, as is often the case in particle physics, the localised solutions on the half-line  found here may be interpreted as extra-dimensional excitations that are localised near a brane by quantum nonlinearities. We also think that the result of Sect.(2) which indicated a nonperturbative quantum vacuum might have implications in cosmology.

It is straighforward to construct some other exact solutions depending on more free parameters, for example by constructing the regularised version of the q-deformed equation in Sect.(3) or by replacing (\ref{deff}) by an appropriate Fourier series.  As for the equations we have considered, we have indicated some avenues for obtaining other localised solutions in Sect.(4). Finally we note that the solutions studied in this paper do not deform continuously to solutions of the linear theory as the nonlinearity vanishes. The perturbative effect of the nonlinearity on the usual solutions of the linear Schrodinger equation are studied elsewhere \cite{gelo}.

\section*{Acknowledement}
We thank an anonymous referee for helpful comments.

\end{document}